\documentclass[12pt,tightenlines,
prd,aps,amssymb,amsmath]{revtex4}
\usepackage{graphics}
\usepackage[pdftex]{graphicx}
\begin{document}


\title{Meta-CTA Trading Strategies and Rational Market Failures}

 \author{ Bernhard K  Meister}

 \affiliation{MK2, Vienna, Austria}
\date{\today }

\begin{abstract}
Investors trade shifting    prices, portfolio values, and
 in turn their ability to borrow. 
  Concentrated ownership,
   high price impact and low collateral requirements  
 are  propitious 
for
  arbitrage. 


\end{abstract}
\thanks{ email: ${\rm bernhard.k.meister@gmail.com 
}$ }

\maketitle



\section{Introduction}
\noindent The paper is the sequel to \cite{meister2016}\footnote{CTA stands for Commodity Trading Advisors, which subsumes a range of investment strategies.}. Previously,  the behaviour of a Kelly-optimising investor in a two asset economy  was discussed. One asset is riskless with interest rate $r$, and the other is  risky with drift $\mu$ and volatility $\sigma$. The Kelly-optimiser starts with a riskless portfolio and then diverts the fraction $(\mu-r)/\sigma^2$ to the risky one\footnote{The Kelly fraction optimises portfolio growth and it is time scale invariant unlike the Sharpe ratio: $(\mu-r)/\sigma\sim 1/\sqrt{T}$.}. The acquisition of the risky asset is assumed to have a price  impact of the form $k A^{\gamma}$ with $A$ the number of units purchased as fraction of the average volume for the transaction time interval, $\gamma$ a coefficient close to one half, and $k$ a constant related to $\sigma$. The price impact function can more generally be written as $f(A)$. 
The wish to keep the optimal fraction in the risky asset entails a further portfolio adjustment, since the price impact shifts the required weights. 
An infinite regress  ensues with the application of the Kelly criterion at each step. Depending on the chosen parameters these modifications can  be divided into three different types. The portfolio adjustments can  incrementally increase, decrease or alternate between positive and negative changes of increasing or decreasing size. This can be compressed by summing up an infinite sum, which can converge or diverge either to infinity or  in the manner of Grandi's series, i.e. $1,-1,1,\dots\,\,$, have no unique outcome. 
 
The Kelly based investors misjudge the market by assuming it to be described by a known geometric Brownian motion. This is erroneous for two reasons. One, the risky asset price is driven solely by price impact, and two, 
 the amount of data required to reasonably accurately estimate the asset drift  is  in any case not available in markets with regular structural upheavals. 
 To avoid this problem the portfolio managers in this paper are not Kelly-optimisers but instead   bent only on profit generation.
 
  In the next section an arbitrage strategy is  discussed for assets with high ownership concentration and  price impact, which is often synonymous with low liquidity. Collateralised borrowing available for example on specialized DeFi platforms is incorporated to show how exploitable bubbles can form. 
 The penultimate section provides a cyclic trading example, and the conclusion rounds of the paper.

\section{ Regret-free Gluttony}
\noindent 
The toy model considered in this section encompasses the same riskless and risky asset as above. The market impact function is also left unchanged.  Novel is that borrowing is only possible up to a percentage $\alpha$, i.e. one minus the haircut, of the portfolio market value.  This seems, especially in the world of automated lending on DeFi platforms, not unreasonable.

In this model, if the risky asset  has sufficiently large price impact, an opportunity arises for a profit-maximizing investor with a suitably large holding  and access to a  lending market, where the risky asset can be provided as collateral and the riskless asset can be borrowed. A  cyclic strategy of borrowing the riskless asset and converting it into the risky asset can be used to build up  successively larger  positions. 
In addition, under the right conditions more than the initial position held by the investor can be diverted from the borrowed funds while still maintaining the upward trajectory of the risky asset price. As a consequence, even the inability to unwind effectively a large position in the risky asset  will only dent but not wipe out the profitability of the strategy. 

In this and the subsequent paragraphs the notation is defined and the discrete as well as continuous versions of the problem formulated. 
The starting position in the risky asset is $x_0$, in the riskless asset is $y_0$ and the borrowed amount is $z_0$, while the exchange rate is $S_0$. The market can be described with the  four component vector 
\begin{eqnarray}
(x_0,y_0,z_0,S_0). \nonumber
\end{eqnarray}
Let's further assume $y_0$ is zero. 
In the first step the borrowed amount is maximised up to the limit $\alpha x_0 S_0$.
The newly borrowed amount is 
\begin{eqnarray}
y_1= \alpha x_0 S_0-z_0=z_1-z_0,\nonumber
\end{eqnarray}
and this leads to 
\begin{eqnarray}
(x_0, y_1, z_0+y_1, S_0).\nonumber
\end{eqnarray}
In the second step the riskless asset is changed to the risky asset 
\begin{eqnarray}
(x_0+y_1/S_1, 0, z_0+y_1, S_1)=(x_1, 0, z_1, S_1)\nonumber
\end{eqnarray}
with the new exchange rate due to the stipulated price impact of 
\begin{eqnarray}
S_1 = S_0 +f \big(y_1 /S_1\big). \nonumber
\end{eqnarray}
A repeat of the first step, i.e. the readjustment of the maximal borrowable amount, leads to
\begin{eqnarray}
(x_1,y_2 , z_2, S_1),\nonumber
\end{eqnarray}
with $y_2=\alpha x_1 S_1-z_1=z_2-z_1$.
The second step with a switch of the newly borrowed riskless asset into the risky one is also repeated. This two step dance continues ad infinitum. In the real world, unlike in the idealised model, something should eventually break and the scheme collapse, since lending pools are finite, price impact is non-constant and unusual price behaviour  of an asset with substantial market capitalisation attracts attention. 

As an aside, the cycles are assumed to be repeated fast enough such that the interest rate  charged on the loan can be neglected in first approximation.
If price impact is non-existent, then the amount an investor can borrow over multiple periods is approximately 
\begin{eqnarray}
\sum_{n=1}^{\infty}\alpha^n x_0 =\frac{\alpha  }{1-\alpha}x_0. \nonumber
\end{eqnarray}
Instead, price impact especially in low liquidity assets is assumed to be significant and permanent, especially when the strategy is implemented in a compressed way such that the market has no time to relax back to earlier levels, as is the case of temporary market impact.
 
The three key relationships expressing the link between $x_j$, $y_j$, $z_j$ and $S_j$ can be  written as 
\begin{eqnarray}
y_{j}=&\alpha(x_j S_j-x_{j-1}S_{j-1})   \,\,\,&(*), \nonumber\\
y_j=&(x_{j+1}-x_{j}) S_j \,\,\,&(**),\nonumber\\
f\big(y_{j+1}/S_{j+1}\big) =&  k\big(y_{j+1}/S_{j+1}\big)^\gamma=S_{j+1}-S_{j}\,\,\,&(***).\nonumber
\end{eqnarray}
Expanding in $\Delta t$ and ignoring higher order terms  the  equations have the form
\begin{eqnarray}
(*) \,\, \&\,\,(**)&:& \,\,\,\,\,\,\,\,\,\alpha\big(x(t) \partial_t S(t) +S(t) \partial_t x(t)\big)\Delta t =S(t)\partial_t x(t) \Delta t, \nonumber \\
(***)&:& \,\,\,\,\,\,\,\,\,\,\,\,f\big(\partial_t x(t)  \,\,\Delta t\,\big)=\partial_tS(t)  \,\,\Delta t.\nonumber
\end{eqnarray}
An Ansatz to solve the above equations in the continuum limit is
\begin{eqnarray}
&\hat{x}(t) &\sim \, t^m,\nonumber\\
&\hat{S}(t) &\sim \, t^n,\nonumber
\end{eqnarray}
for the impact function given by 
\begin{eqnarray}
f\big(\partial_t x(t ) \,\,\Delta t\,\big)= \hat{k}(\partial_t x(t))^{\gamma} \,\, \Delta t,\nonumber
\end{eqnarray}
and $k=\hat{k} (\Delta t)^{1-\gamma}$ to obtain   $m=\alpha (1-\gamma)/(\alpha-\gamma(1 -\alpha )) $ and $n =(1-\alpha)(1-\gamma)/(\alpha-\gamma(1 -\alpha ))$.  For the popular square root impact, i.e. $\gamma=1/2$, the polynomial growth coefficients simplify to $n=(1-\alpha )/(3 \alpha - 1 )$ and $m=\alpha/(3 \alpha - 1 )$.
If reasonable initial conditions at $t=1$ are chosen, then the polynomial increase of the   amount held $x(t)$ and the price of the risky asset $S(t)$  follows.
The instability is replicated 
in the   discrete version of the equations.

As the borrowed amount increases with each round the toy model under suitable conditions 
results in an exploitable instability, since  part of the newly borrowed money can be extracted round by round instead of being reinvested, while still maintaining a 
 rise in the risky asset price. 
If done repeatedly, then  more can be extracted than the initial  investment. When eventually the growth of the asset is interrupted through some extraneous event the speculator is richer and the lender has to scramble to unwind collateral of dubious value. This is  dubbed a `rational market failure'.

A possible example is the fracas this June associated with the lending platform Solend linked with Solana, where the possible negative consequences of having to forcibly unwind a  single large borrower was sufficiently daunting to lead to a temporary change in the rules. 
A similar sentiment is reflected in a statement attributed to Keynes, `owe your banker \pounds 1,000 and you are at his mercy; owe him \pounds 1 million and the position is reversed'.
 The same maybe applies to aggressive hedge fund strategies, as exemplified by Archegos Capital Management, which collapsed in a frenzy in early 2021.  Archegos' highly leveraged trading strategy as reported by the media seems  inexplicable, since unwinding concentrated positions in a small number of stocks normally entails significant mark-to-market losses. If instead excess collateral can be extracted, i.e. a hypothetical hedge fund is able to syphon of capital, then maybe only the lenders are left with losses.  
 It is not suggested that this is necessarily what happened at Archegos Capital, since published reports are   fragmentary and require careful study, which the author has not undertaken. 

  Long-Term Capital Management L.P. 
  from the mid-nineties  aimed  more intriguingly 
  at convergence instead of divergence, and the fund's exceptional ability to accumulate leverage in conjunction with other market participants was employed for a time successfully.
  Convergence 
  became  fashionable 
  leading to a decline of spreads, relative volatility and price impact. Relative value positions soared in popularity and size,  
   until the market had, to put it figuratively, a `Wile E Coyote moment'. 
A  strategy based on roundtrip trading is described next.
\section{ Roundtrip Trading} 
\noindent
Not only accumulative but also profitable roundtrip trading strategies can be designed. A qualitative summary of such a strategy is next given. If permanent price impact varies in a predictable way, then purchasing assets at times of high price impact and selling the same amount at times of low price impact is associated in isolation with loses, but can in conjunction with a substantial long position in the underlying be profitable, if the increase of the value of the underlying position more than off-sets the loses of the roundtrip component\footnote{The inverse strategy of selling in times of high impact and buying in times of low impact, while maintaining a large short position, can also be attractive.}.

Assume the position held in the risky asset by the arbitrageur is $N$ units, the number of units involved in the round-trip trade are $N_r$ and the round-trip cost per unit of asset to be $\Delta C$, which is a combination of transaction cost, bid-offer spread and cost due to temporary and permanent price impact. The permanent price impact is taken to be $\Delta S$. The possibility that trades can have permanent as well as transitory price impact is discussed in the literature and is here taken as a given. 
Roundtrip trading is viable, if $N \Delta S\geq N_r \Delta C$, and can be repeated to magnify the result.

 Boosterism  due to this kind of price steering is particularly seductive in crypto-finance, where intellectual as well as physical capital  inflow and survival depend  on maintaining a reasonable  position in the market capitalisation ranking.
Additional material  can be found in \cite{meisterprice2022} and references therein. A short conclusion follows.

\section{Conclusion}
\noindent  
In the toy model of the original Meta-CTA paper\cite{meister2016} the unintended consequence  of an investor conscientiously applying the Kelly criterion could be the careering of the asset price to infinity.  In this paper what was earlier considered a `bug' is turned into the 
`feature'.

For a market with aligned low liquidity, high ownership concentration and low collateral requirement for borrowing the price can become detached from  any conceivable value. 
For tokens in cryptofinance maybe both the ability and even the necessity to carry out   boosterism   to stay in the game exists. 
 The usual caveats apply: the world is complicated and it is easier to hypothesize than to prove. 
 
A natural way to mitigate excesses in the paper's toy model  would be to link the haircut, i.e $1-\alpha$, to the position size of the largest borrowers and available liquidity. This requires non-trivial adjustments to the risk management formulae encoded in DeFi platforms\footnote{As a comparison, centralised exchanges   adjust depending on volatility initial and variation margin.
}.  
 

 Shakespeare has 
Polonius in Hamlet proclaim, `Neither a borrower nor a lender be, For loan oft loses both itself and friend, And borrowing dulls the edge of husbandry'.
Leverage, as an accelerant, underpins the capitalist economy and Polonius' platitude is almost universally  dismissed. Nevertheless, once the genii of easy money is out of the bottle enticing it back requires guile. 
 Felix Somary a Viennese born banker nicknamed `the raven of Zurich'  suggested at various points during the turbulent first half of the 20th century that short sharp defaults are preferable to long-drawn out and ultimately futile evasive actions. 
His insights, beneficial for 
the  private bank he led,  were otherwise largely ignored. 

\noindent
Discussions with D\,C\,Brody and H\,C\,W\,Price are gratefully acknowledged.


\nocite{*}

\end{document}